\documentclass[fleqn,usenatbib]{mnras}

\usepackage{newtxtext,newtxmath}

\usepackage[T1]{fontenc}

\DeclareRobustCommand{\VAN}[3]{#2}
\let\VANthebibliography\thebibliography
\def\thebibliography{\DeclareRobustCommand{\VAN}[3]{##3}\VANthebibliography}

\usepackage{graphicx}	
\usepackage{amsmath}	

\newcommand{\masy}{mas\,yr$^{-1}$}
\newcommand{\feh}{[Fe/H]}
\newcommand{\teff}{$\mathrm{T_{eff}}$}
\newcommand{\logg}{$\log g$}
\newcommand{\kms}{km s\textsuperscript{-1}}
\newcommand{\ms}{m s\textsuperscript{-1}}
\newcommand{\Rjup}{R\textsubscript{Jup}}
\newcommand{\Mjup}{M\textsubscript{Jup}}
\newcommand{\Msun}{M\textsubscript{$\odot$}}
\newcommand{\Rsun}{R\textsubscript{$\odot$}}

\newcommand{\Nstar}{NGTS-19}
\newcommand{\Nstarb}{NGTS-19b}

\newcommand{\Rsuma}{$0.0359 ^{+0.0015}_{-0.0013}$}
\newcommand{\RRatio}{$0.1182 ^{+0.0037}_{-0.0035}$}
\newcommand{\cosinc}{$0.0223 ^{+0.0018}_{-0.0019}$}
\newcommand{\impact}{$0.730 ^{+0.033}_{-0.041}$}
\newcommand{\Epoch}{$2458533.0207 ^{+0.0011}_{-0.0011}$}
\newcommand{\Period}{$17.839654 ^{+0.000037}_{-0.000038}$}
\newcommand{\Periodshort}{$17.84$}

\newcommand{\eccone}{$0.5322 ^{+0.0042}_{-0.0042}$}
\newcommand{\ecctwo}{$-0.306 ^{+0.014}_{-0.014}$}
\newcommand{\sysvel}{$-33.9166 ^{+0.0097}_{-0.0105}$}
\newcommand{\rvamp}{$6.492 ^{+0.063}_{-0.062}$}

\newcommand{\ldcngtsone}{$0.64 ^{+0.22}_{-0.24}$}
\newcommand{\ldcngtstwo}{$0.53 ^{+0.27}_{-0.29}$}
\newcommand{\ldctessone}{$0.37 ^{+0.25}_{-0.18}$}
\newcommand{\ldctesstwo}{$0.39 ^{+0.32}_{-0.25}$}
\newcommand{\ldcsaaoone}{$0.37 ^{+0.29}_{-0.21}$}
\newcommand{\ldcsaaotwo}{$0.45 ^{+0.31}_{-0.28}$}

\newcommand{\BDMass}{$69.5 ^{+5.7}_{-5.4}$}
\newcommand{\BDRad}{$1.034 ^{+0.055}_{-0.053}$}
\newcommand{\BDDens}{$77 ^{+17}_{-13}$}
\newcommand{\SMAAU}{$0.1296 ^{+0.0074}_{-0.0072}$}
\newcommand{\SMAscaled}{$27.9 ^{+1.6}_{-1.6}$}
\newcommand{\ecc}{$0.3767 ^{+0.0061}_{-0.0061}$}
\newcommand{\periastron}{$330.1 ^{+1.4}_{-1.3}$}
\newcommand{\incl}{$88.72 ^{+0.11}_{-0.11}$}
\newcommand{\eqtemp}{$543^{+17}_{-16}$}
\newcommand{\transitwidth}{$4.252 ^{+0.077}_{-0.074}$}

\newcommand{\secphase}{$0.7075 ^{+0.0055}_{-0.0055}$}

\newcommand{\vsini}{$2.1 \pm 0.4$}

\newcommand{\RMAmp}{$13.4 ^{+2.7}_{-2.6}$}

\title[High mass transiting brown dwarf]{NGTS-19b : A high mass transiting brown dwarf in a 17-day eccentric orbit}

\author[J. S. Acton et al.]{
\parbox{\textwidth}{
Jack S.~Acton,$^{1}$\thanks{E-mail:ja466@le.ac.uk} 
Michael R.~Goad,$^{1}$
Matthew R.~Burleigh,$^{1}$
Sarah L.~Casewell,$^{1}$
Hannes~Breytenbach,$^{2,3}$
Louise D.~Nielsen,$^{4}$
Gareth Smith,$^{5}$
David~R.~Anderson,$^{6,7}$
Matthew P. Battley,$^{6,7}$
Daniel Bayliss,$^{6,7}$
Fran\c{c}ois Bouchy,$^{4}$
Edward M.~Bryant,$^{6,7}$
Szil\'ard~Csizmadia$^{8}$
Philipp Eigm\"uller,$^{8}$
Samuel Gill,$^{6,7}$
Edward~Gillen,$^{9,5}$\thanks{Winton Fellow}
Nolan~Grieves,$^{4}$
Maximilian N. G{\"u}nther,$^{10}$\thanks{Juan Carlos Torres Fellow}
Beth A. Henderson,$^{1}$
Simon T. Hodgkin,$^{11}$
James A. G. Jackman,$^{12,6,7}$
James S. Jenkins,$^{13,14}$
Monika~Lendl,$^{4}$
James~McCormac,$^{6,7}$
Maximiliano Moyano,$^{15}$
Richard P. Nelson,$^{9}$
Ramotholo~R.~Sefako,$^{2}$
Alexis~M.~S.~Smith,$^{8}$
Manu~Stalport,$^{4}$
Jessymol~K.~Thomas,$^{2}$
Rosanna H.~Tilbrook,$^{1}$
St\'{e}phane~Udry,$^{4}$
Richard~G.~West,$^{6,7}$
Peter~J.~Wheatley,$^{6,7}$
Hannah~L.~Worters,$^{2}$
Jose I.~Vines,$^{13}$
Douglas R. ~Alves$^{13}$
}
\\
$^{1}$School of Physics and Astronomy, University of Leicester, University Road, Leicester, LE1 7RH, UK\\
$^{2}$ South African Astronomical Observatory, P.O Box 9, Observatory 7935, Cape Town, South Africa\\
$^{3}$ Department of Astronomy, University of Cape Town, Rondebosch 7700, Cape Town, South Africa\\
$^{4}$Observatoire de Gen{\`e}ve, Universit{\'e} de Gen{\`e}ve, 51 Ch. des Maillettes, 1290 Sauverny, Switzerland\\
$^{5}$Astrophysics Group, Cavendish Laboratory, J.J. Thomson Avenue, Cambridge CB3 0HE, UK\\
$^{6}$Centre for Exoplanets and Habitability, University of Warwick, Gibbet Hill Road, Coventry CV4 7AL, UK\\
$^{7}$Dept.\ of Physics, University of Warwick, Gibbet Hill Road, Coventry CV4 7AL, UK\\
$^{8}$Institute of Planetary Research, German Aerospace Center, Rutherfordstrasse 2., 12489 Berlin, Germany\\
$^{9}$Astronomy Unit, Queen Mary University of London, Mile End Road, London E1 4NS, UK\\
$^{10}$Department of Physics, and Kavli Institute for Astrophysics and Space Research, Massachusetts Institute of Technology, Cambridge, MA 02139, USA\\
$^{11}$Institute of Astronomy, Madingley Road, Cambridge, CB3 0HA, UK\\
$^{12}$School of Earth and Space Exploration, Arizona State University, Tempe, AZ 85287\\
$^{13}$Departamento de Astronomia, Universidad de Chile, Casilla 36-D, Santiago, Chile\\
$^{14}$Centro de Astrof\'isica y Tecnolog\'ias Afines (CATA), Casilla 36-D, Santiago, Chile\\
$^{15}$Instituto de Astronom\'ia, Universidad Cat\'olica del Norte, Angamos 0610, 1270709, Antofagasta, Chile
}

\date{Accepted XXX. Received YYY; in original form ZZZ}

\pubyear{2020}

\begin{document}
\label{firstpage}
\pagerange{\pageref{firstpage}--\pageref{lastpage}}
\maketitle

\begin{abstract}
We present the discovery of \Nstarb\, a high mass transiting brown dwarf discovered by the Next Generation Transit Survey (NGTS). We investigate the system using follow up photometry from the South African Astronomical Observatory, as well as sector 11 TESS data, in combination with radial velocity measurements from the CORALIE spectrograph to precisely characterise the system. We find that \Nstarb\ is a brown dwarf companion to a K-star, with a mass of \BDMass\, \Mjup\ and radius of \BDRad \Rjup. The system has a reasonably long period of \Periodshort\ days, and a high degree of eccentricity of \ecc. The mass and radius of the brown dwarf imply an age of $0.46 ^{+0.26}_{-0.15}$ Gyr, however this is inconsistent with the age determined from the host star SED, suggesting that the brown dwarf may be inflated. This is unusual given that its large mass and relatively low levels of irradiation would make it much harder to inflate. \Nstarb\ adds to the small, but growing number of brown dwarfs transiting main sequence stars, and is a valuable addition as we begin to populate the so called brown dwarf desert.
\end{abstract}




\section{Introduction}

Brown dwarfs are substellar mass objects which bridge the gap between planets and stars. These are are objects with radii similar to that of Jupiter, but with masses ranging between 13 \Mjup\ and  $\sim$80 \Mjup\ (\citealt{Spiegel2011, Baraffe2002}). The lower mass limit corresponds to the minimum mass at which deuterium burning can occur, below which lie the planets. Whilst the upper mass limit is the classical hydrogen burning limit, above which objects are considered to be low mass stars. 

The first unambiguous detections of brown dwarfs were the discoveries of Gilese 229B \citep{Nakajima1995} and Teide 1 \citep{Rebolo1995}. Since then there have been thousands of brown dwarfs discovered, the vast majority of which are isolated objects discovered by wide field photometric surveys (e.g \citealt{Pinfield2008, Folkes2012, Reyle2018, Schneider2020, Kirkpatrick2020}). This is a result of the fact that brown dwarfs cool as they age, due to a lack of nuclear fusion in their cores, making them much easier to detect at the long wavelengths that these surveys typically operate (e.g WISE; \citealt{Wright2010}). 

Not all brown dwarfs exist in isolated systems. Many have been discovered as companions to main sequence stars, a large number of which have been identified via direct imaging (\citealt{Nielsen2019, Vigan2020}). However, brown dwarfs are much fainter than stars, and thus we can only resolve them at large orbital separations. To find close in brown dwarfs, we need to look at the effect they have on their host star (e.g., via transits or radial velocity measurements). 

In recent years there have been an increasing number of transiting brown dwarfs discovered around main and pre-main sequence stars by exoplanet surveys (e.g \citealt{Csizmadia2015,Bayliss2017, Jackman2019}). Given their Jovian sized radii, we would expect these objects to be easy to detect given the vast number of known hot Jupiters around main sequence stars. Similarly, their large masses result in easy to detect radial velocity shifts (\kms\ scale rather than \ms\ for planetary mass objects) which should also aid in discovery. However despite this, there are just 28 known transiting brown dwarfs around main sequence stars (\citealt{Carmichael2020,Palle2021}), an unusually small number given the fact there are over 4000 known exoplanets\footnote{According to the NASA Exoplanet Archive, March 2021}.



This phenomenon is known as the brown dwarf desert (e.g. \citealt{Grether2006}) and was identified as being a dearth of brown dwarfs orbiting main sequence stars within 3~AU, in contrast to the large number of binary stars with close orbits \citep{Marcy2000}. This was further highlighted by the discovery of large numbers of exoplanets in short period orbits ($\sim$days), but very few brown dwarfs  \citep{Cumming2008}. Recent work with SuperWASP confirmed that this desert still remains \citep{Triaud2017}, despite the vast number of transiting objects discovered by $Kepler$ \citep{Borucki2010} and $TESS$ \citep{Ricker2015}. The desert is thought to be a result of the differences between formation mechanisms for planets and brown dwarfs, although recent transiting brown dwarf discoveries have called into question the nature of this so called desert \citep{Carmichael2019}.


Due to their scarcity, it is important that we understand and characterise those transiting brown dwarfs that have been discovered, in particular their masses and radii. Here, transiting brown dwarfs around main sequence stars provide us with an advantage over isolated field brown dwarfs. Their radii can be measured from the depth of their transit in the lightcurve of their host star, whilst their masses can be measured using radial velocity measurements. Provided the host star parameters are well defined, this allows for accurate determination of these fundamental parameters.

These parameters are particularly valuable when combined with an accurate measurement of system age. As their masses are too small to fuse hydrogen, brown dwarfs cool and undergo gravitational contraction as they age. This contraction occurs most quickly up to an age of 1 Gyr and gradually decreases with time (\citealt{Baraffe2003, Saumon2008, Burrows2011, Phillips2020}). With knowledge of the age we can compare the radius of the brown dwarf to that expected by models of gravitational contraction (e.g \citealt{Phillips2020}). Equally, we can use a well defined mass and radius measurement to estimate the age of the brown dwarf, by comparing it with stellar isochrones of a variety of ages to see which provides the best fit.

However whilst there are a large number of known brown dwarfs in stellar clusters with well defined ages \citep{Pearson2020}, this is not the case for transiting brown dwarfs. There are are very few with accurate age measurements as a result of cluster association (e.g., \citealt{Gillen2017, Beatty2018, David2019}). So the age often needs to be inferred by some other means for example through fitting the spectral energy distribution (SED) of the host star \citep{Choi2016}, or using stellar evolutionary tracks. However this can result in uncertainties of up to a few Gyrs in some cases (e.g \citealt{Bayliss2017, Nowak2017}).

Age measurements are particularly important in the case of transiting brown dwarfs, as we are able to compare the brown dwarf masses and radii to evolutionary models (e.g \citealt{Baraffe2003, Marley2018}). This is one of the best ways to compare the accuracy of these models, as we are comparing them to brown dwarfs whose properties have been directly measured from their interaction with their host stars, allowing for an improved understanding of how these objects evolve as they age. This is particularly important at the high mass (60--80 \Mjup) end of the distribution, where \cite{Baraffe2003} predict the largest changes in radii with age. 

It is well established that low mass stars in eclipsing binaries show significant scatter in the relationship between mass, radius and luminosity when compared with evolutionary models \citep{Parsons2018}. However with so few transiting brown dwarfs yet discovered, whether this relationship continues into the substellar regime is unclear. It has been suggested that brown dwarfs may be able to be inflated if they orbit host stars that are particularly active \citep{Casewell2020}. A comparison with evolutionary models would suggest that high mass brown dwarfs are much less likely to be inflated, regardless of the effect of their host star. Additionally, detailed investigation of individual systems has shown that stellar irradiation alone cannot explain discrepancies when they are observed \citep{Beatty2018}. However with so few transiting brown dwarfs known it is challenging to make any clear determination of the cause of these discrepancies. Hence it is important that we discover and characterise as many of these systems as possible. 



In this paper we present the discovery of \Nstarb\, a high mass brown dwarf in an eccentric 17-day orbit around a main sequence K-star.  We make use of high precision photometric and spectroscopic follow up to derive an accurate mass and radius for the brown dwarf. We then place this system in the context of the growing number of brown dwarfs being identified around main sequence stars, and explain how this, and future discoveries will aid in understanding the formation and evolution of these systems.

\section{Observations}

\begin{table*}
	\centering
	\caption{Summary of photometric and spectroscopic observations of \Nstar.}
	\label{tab:obs_summary}
	\begin{tabular}{ccccccc} 
    \hline
Observation type & Telescope & Band  & Cadence & Total integration time & Period & Notes\\
\hline
Photometry	& NGTS	&	520-890\,nm	&	13\,s   & 148\,nights	    &	26/01/17-17/09/17 & 3 Transits\\
Photometry	& SAAO	&	V	        &	60\,s   & 400\,mins 	    &	19/07/20 & Single Observation\\
Photometry	& TESS	&  600-1000\,nm	&	1800\,s & 28\,days	    &	22/04/19-20/05/19 & 2 Transits\\
Spectroscopy& CORALIE	&	390-680\,nm	        &  45\,mins &	    4.5\,hours	    &  05/02/20-27/02/21          & Eight RVs \\

	\hline
    \end{tabular}
\end{table*}

\begin{table}
	\centering
	\caption{Stellar properties and colour magnitudes for \Nstar\, from 2MASS \citep{Skrutskie2006}, Gaia \citep{GAIA2018}, TicV8 \citep{Stassun2019} and NGTS \citep{Wheatley2018}.}
	\label{tab:stellar}
	\begin{tabular}{ccc} 
	Property	&	Value		&Source\\
	\hline
	Gaia I.D.		&	DR2 6226795997504049664	& Gaia \\
    TIC I.D.		&	48481940	& TIC v8 \\
    R.A. (J2000)		&	15:16:31.6				&	NGTS	\\
	Dec	 (J2000)		&	-25:42:17.24	&		  NGTS \\
    $\mu_{\alpha}$ (\masy)& $-45.796 \pm0.054$  & Gaia\\
    $\mu_{\delta}$ (\masy) & $-14.746 \pm0.038$ & Gaia\\
    Parallax (mas) & $2.6666 \pm0.0285$ & Gaia\\
    $G$			&$13.83$	&Gaia\\
    NGTS		&$13.20$				& NGTS\\
    TESS		&$13.20$				& TIC v8\\
    $B$	    	&$15.18$				& TIC v8\\
    $V$ 		&$14.12$				& 2MASS\\
    $J$			&$12.27$		&2MASS	\\
   	$H$			&$11.82$		&2MASS	\\
	$K$			&$11.70$		&2MASS	\\
    
	\hline
	\end{tabular}
\end{table}

\Nstar\ was initially discovered using photometry from the Next Generation Transit Survey (hereafter NGTS;  \citep{Wheatley2018}). Follow-up observations were performed with the Sutherland High Speed Optical Cameras (SHOC) \citep{Coppejans2013} on the South African Astronomical Observatory (SAAO) 1-m telescope. This photometry was then used in conjunction with observations from the Transiting Exoplanet Survey Satellite (TESS, \citealt{Ricker2015}). We obtained high resolution spectra with the CORALIE Spectrograph \citep{Queloz2000} to determine the mass of the companion. These observations are detailed in Table \ref{tab:obs_summary} and described below.

\subsection{NGTS Photometry}\label{ngts}

Transits of \Nstarb\ were initially detected in survey photometry from the Next Generation Transit Survey  (\citealt{Wheatley2018}). NGTS is a wide-field ground based survey for transiting exoplanets operating at ESO's Paranal observatory in Chile. It consists of an array of 12 fully automated 20~cm telescopes which operate independently to survey large areas of the sky each night. NGTS is optimised for observations of K- and M-type stars, with a custom bandpass range of 520 to 890-nm. NGTS has a wide field of view (instantaneously covering 96 sq deg) and delivers high cadence (every $\sim$ 13 seconds) photometry with high precision (1 mmag per hour for an $I$=14 magnitude star).

\begin{figure}
   \centering
	\includegraphics[width=\columnwidth]{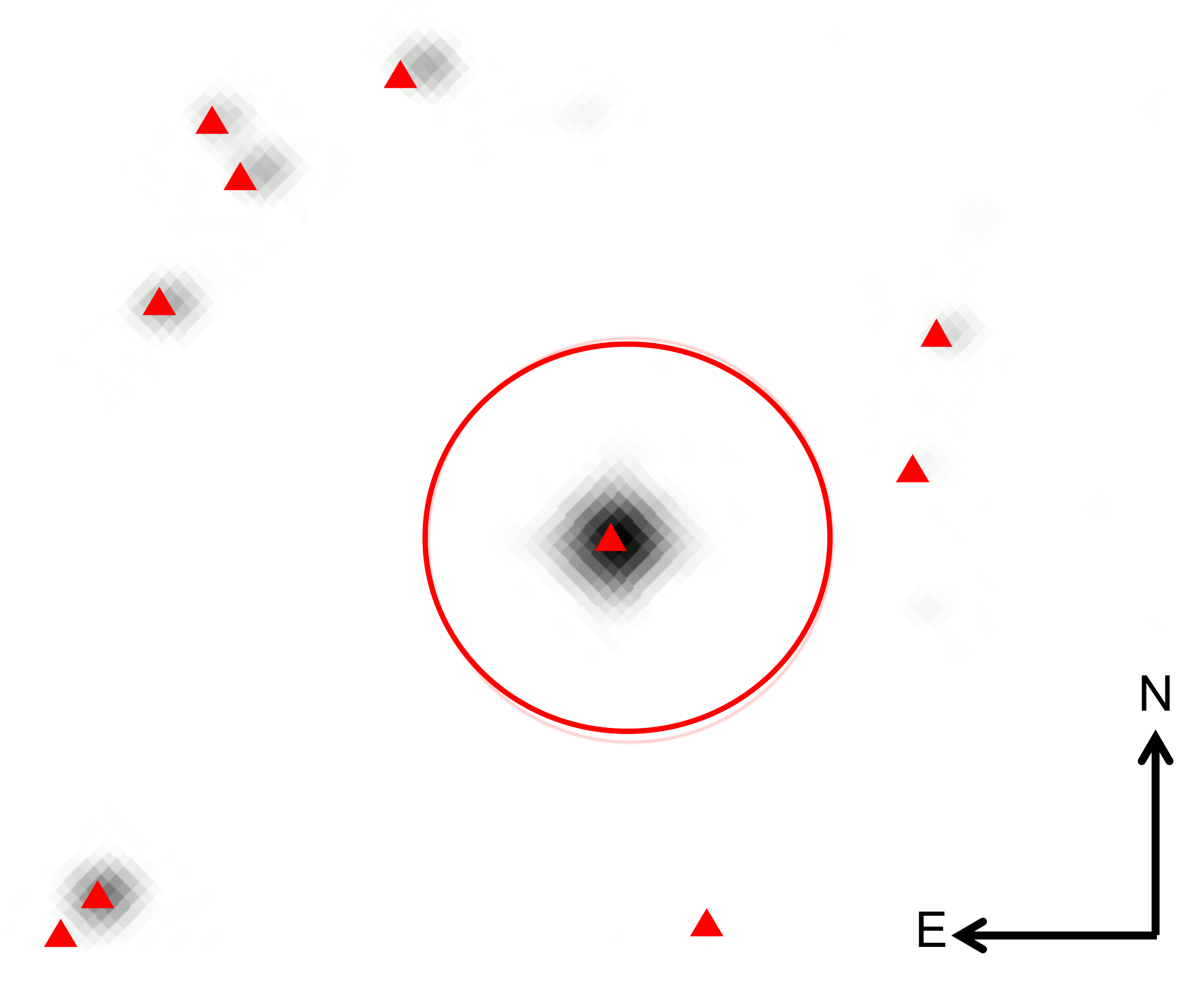}
   \caption{Digital Sky Survey (DSS) image of \Nstar. The red triangles indicate the positions of nearby objects that are identified in Gaia DR2. The red circle shows the NGTS aperture used to create the NGTS lightcurve. \Nstar\ is the only known source within the aperture used for reduction.}
   \label{fig:dss}
\end{figure}

\Nstar\ was observed during the 2017 NGTS observing season. The field containing the system (NG1518-2518) was observed for 148 nights between 2017 January 26\textsuperscript{th} and 2017 July 17\textsuperscript{th}, and in total we obtained $200,547$ science images at a cadence of 13 seconds. 
The star is shown in Figure \ref{fig:dss}. The star is well isolated, and there are no additional Gaia DR2 sources in the NGTS aperture that could dilute the depth of the transit. 
We also note that none of the nearby Gaia DR2 sources have a parallax or proper motion that is consistent with being physically associated with \Nstar. The magnitudes of the system in various bandpasses, as well as positional information, is provided in Table \ref{tab:stellar}. 

\begin{figure*}
   \centering
	\includegraphics[width=\textwidth]{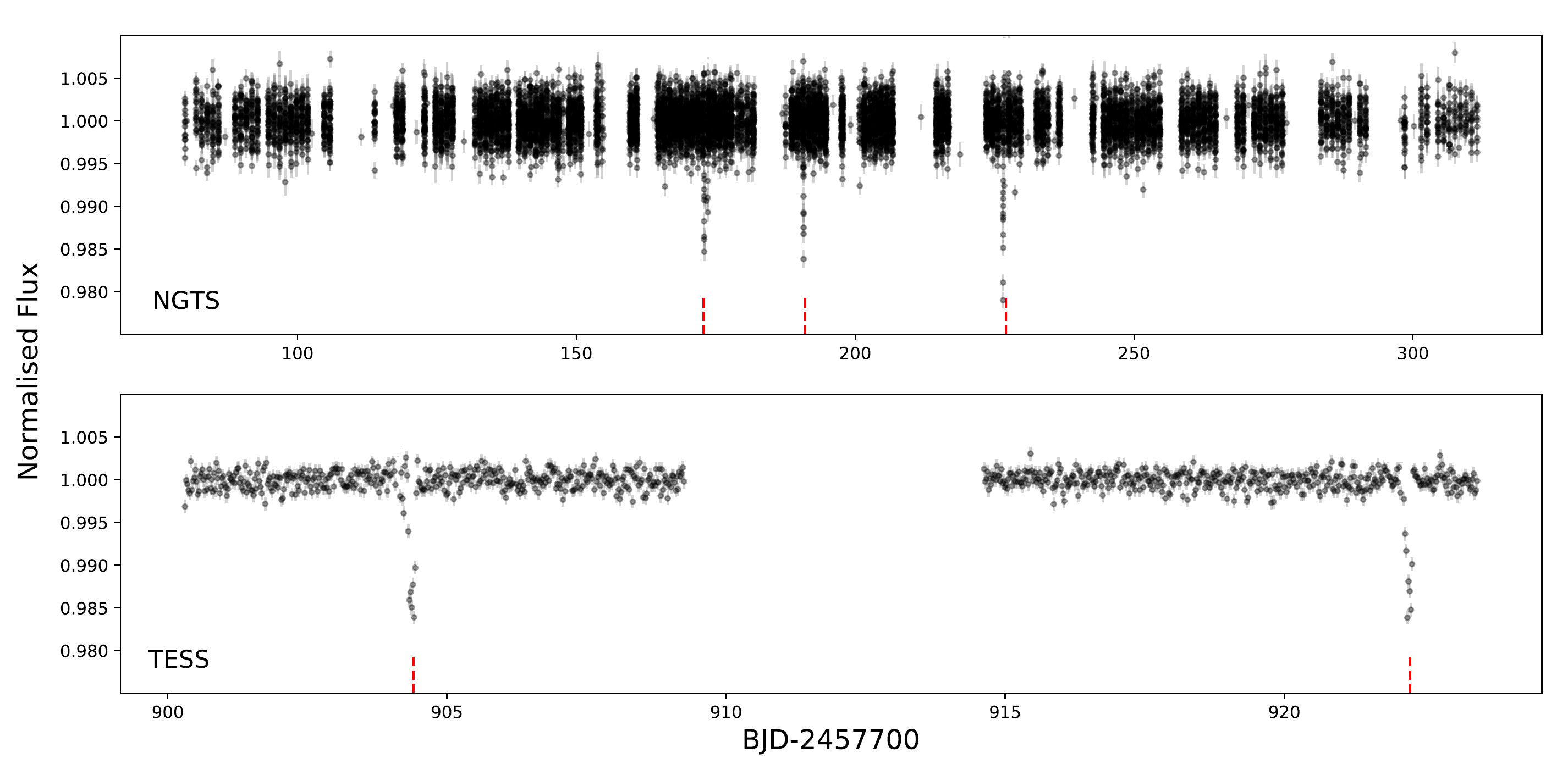}
   \caption{Upper: NGTS Lightcurve of \Nstar\ comprising of 148 nights of observations, binned to 5 minutes. NGTS detects three transits of the system, indicated by the red lines. Lower :TESS Lightcurve for \Nstar\ obtained during sector 11 of the mission (TIC-48419840), observed at 30 minute cadence in full frame images. Two transits are seen during the sector, spaced approximately 17.8 days apart, and are indicated by the red dashed line.}
   \label{fig:LCS}
\end{figure*}

The raw lightcurve was cleaned using an implementation of the \textsc{SysRem} algorithm \citep{Tamuz005}. Transits were then detected using the \textsc{orion} algorithm, which is a custom implementation of the usual Boxed Least Squares (BLS; \citealt{Kovacs2016}) algorithm (see e.g., \citealt{Wheatley2018} for more information). 

The NGTS observations captured three transits of the system (see Figure \ref{fig:LCS}). From this, \textsc{orion} detected an orbital period of 17.8 days and transit depth of around 1.5 per cent, corresponding to an object with a radius approximately that of Jupiter. Any shorter period alias could be ruled out by the lack of detection on other nights the star was observed. NGTS candidates are also vetted by a convolutional neural network (CNN) designed to distinguish between transiting signals and false positives  \citep{Chaushev2019}. \Nstar\ received a CNN probability of 0.96, strongly suggesting the detection was from a transiting planetary-sized object. This information gave us good confidence that the signal was real, and we decided to undertake follow-up observations.

\subsection{TESS Photometry}
\Nstar\ was observed by TESS in Sector~11 of the primary mission (TIC-48481940, T=13.2 mag). The system was observed between 2019 April 22\textsuperscript{nd} and 2019 May 20\textsuperscript{th}, in the full frame images at 30 minute cadence. TESS observed \Nstar\ with CCD~3 of camera~1. 

To extract the lightcurve of \Nstar\ from the full-frame images we used a bespoke process, described in more detail in \citet{Gill2020}. In short, we use a custom aperture that is selected based on a flux threshold. Background pixels were selected using an iterative sigma clipping process, and pixels where the median counts exceeded 100 times the standard deviation in the background were identified as the source.  We then used a floating median to identify and mask out systematic flux drops due to spacecraft effects.

TESS detects two transits of \Nstar\, spaced approximately 17-days apart (see Figure \ref{fig:LCS}). It is worth noting that the TESS observations alone cannot rule out the orbital period of the system being half of this, as a third transit would fall in the data gap in the middle of the sector during spacecraft downlink. 
However, this potential shorter period is not compatible with either the NGTS photometry (Section \ref{ngts}) nor the radial velocity measurements (Section \ref{RV}), which are both compatible with a 17.8~day period,  thus highlighting the need for additional data to interpret the TESS photometry. We note that \Nstar\ will be re-observed in TESS sector 38 of the extended mission. Based on our established ephemeris TESS should then observe a further two transits.

\subsection{SAAO Photometry}
Due to the period of this system, the combined photometry of NGTS and TESS contained only a small number of transits. Thus we obtained additional photometry of \Nstar\ using the 1-m telescope at SAAO with the SHOC instrument on 2020 July 19\textsuperscript{th}. This would allow us to obtain more precise measurements of the transit depth and width, and thus increase our precision on the radius of the companion. We observed the transit in the $V$ band in order to check for any colour dependent depth difference which may be indicative of a stellar companion. The observation consisted for $400 \times 60$-sec exposures for a total observation time of  6~hours 40~minutes. 

The data were bias and flat field corrected via the standard procedure, using the \textsc{SAFPhot} Python package\footnote{https://github.com/apchsh/SAFPhot}. \textsc{SAFPhot} was also used to carry out differential photometry, by extracting aperture photometry from the target as well as comparison stars using the 'SEP' package \citep{Barbary2016}. SEP also measured and subtracted the sky background, adopting a box size and filter width which minimised the background residuals measured across the frame after the stars had been masked out. Two comparison stars were used to perform differential photometry on the target, with a 3.8 pixel radius aperture selected to maximise the signal-to-noise.

The observation clearly detects the transit, which occurred almost exactly as predicted by the ephemeris from the NGTS and TESS observations. We also note no significant difference in depth between this lightcurve and the NGTS and TESS observations, despite the fact the SAAO data were obtained using a much bluer filter. This adds further confidence to the companion being sub-stellar in nature. 

\subsection{CORALIE Radial Velocities}\label{RV}
\begin{table}
	\centering
	\caption{Radial Velocities for \Nstar\ obtained with the CORALIE spectrograph}
	\label{tab:RV_summary}
	\resizebox{\columnwidth}{!}{
	\begin{tabular}{ccccccc} 
    \hline
BJD$_\mathrm{TDB}$			&	RV		&RV error &	FWHM& 	Contrast& Bisector Span\\
(-2,450,000)	& (km/s)& (km/s)& (km/s)&(\%) & & \\
		\hline
8884.876	& -30.53  &	0.17&	9.68&	53.50 & -0.097\\
8885.832	& -26.50  &	0.13&	9.57&  53.28 & -0.239\\
8903.859  & -26.15  &	0.14&	9.43&	54.37 & 0.057\\
8911.851  & -36.73  &	0.11&	9.82&	49.47 & -0.299\\
8925.753   & -32.27   & 0.11&  9.15&  52.78 & 0.0212 \\
8928.730   & -36.14   &   0.15&  9.01&  54.19 & -0.014\\ 
9261.805	&-25.77	 &0.14 	 & 9.25	 &58.78	&-0.17\\
9272.782	&-38.40	 & 0.16	& 9.14	 &52.94	&-0.28	\\ \hline

	\end{tabular}
	}
\end{table}

To determine the mass and orbital eccentricity of \Nstarb\ we obtained spectroscopic observations using the CORALIE spectrograph mounted on the Swiss 1.2-metre Leonhard Euler Telescope at ESO's La Silla Observatory, Chile. CORALIE is an echelle spectrograph fed by a 2 arcsec science fibre, capable of 3\ms\ RV precision on bright stars (see e.g \citet{2019A&A...625A..71R} for recent results). As \Nstar\ is relatively faint for a telescope of this size (V = 14.12), we used long (45 minute) exposures to maximise the signal-to-noise. This allowed for precision on a scale of around 100\,m s$^{-1}$,  which is sufficient to distinguish between stellar and sub-stellar companions.

We obtained a total of eight spectra of \Nstar. These were cross correlated with a binary G2 mask using the standard CORALIE pipeline, while discarding the first 20 spectral orders where the signal-to-noise ratio was less than one. The radial velocity associated with \Nstar\, was extracted from the Cross Correlation Function (CCF) by fitting a Gaussian function. The radial velocity measurements show a large amplitude variation ($\sim$ 6.5 \kms) in phase with the period defined by the NGTS photometry, with a significant level of eccentricity. This suggests the presence of a high mass substellar companion. The full radial velocity measurements are shown in Table \ref{tab:RV_summary}. We note that one of the points was obtained close to phase 0, however we ensured that this measurement was not taken during the transit of the brown dwarf.



\section{Analysis}
\begin{table}
	\centering
	\caption{\Nstar\ stellar parameters derived using \textsc{
	specmatch-emp} and \textsc{ariadne}. The \textsc{ariadne} parameters are used for the global modelling of the system in Section \ref{modelling}.}
	\label{tab:stellar_params}
	\renewcommand{\arraystretch}{1.5}
\begin{tabular}{ccc}
\hline
     Parameter &              {\sc specmatch-emp} &   {\sc ariadne}            \\ \hline
     Teff (K) &  4500 $\pm$  110 &  4716$^{+39}_{-28}$    \\
     Log g (cm s\textsuperscript{-2})&  4.62 $\pm$0.12 &  4.571 $^{+0.102}_{-0.093}$  \\
     Radius (\Rsun)&  0.71 $\pm$0.10 &   0.896 $^{+0.040}_{-0.035}$ \\
     \feh &  0.09 $\pm$0.09 & 0.11 $^{+0.074}_{-0.070}$ \\
     Mass (\Msun)&  0.73 $\pm$0.08 &  0.807   $^{+0.038}_{-0.043}$ \\
     Age~(Gyr) &  ----  &  8.5$^{+3.2}_{-6.0 }$ \\
     Distance~(pc) & ---- &  371$^{+15}_{-12}$  \\
     \hline
\end{tabular}
\end{table}

\subsection{Spectral Analysis}\label{specmatch}
We used the CORALIE spectra of \Nstar\ to obtain  initial parameters for the system.  The CORALIE spectra were shifted in wavelength and co-added to create a single high signal-to-noise spectrum for spectral analysis. However, owing to the faintness of the system this combined spectrum still had a relatively low signal-to-noise ratio (8.28). Nonetheless it was sufficient to obtain some initial system parameters that would be refined in the subsequent analysis.

We analysed this stacked spectrum using the template matching code \textsc{specmatch-emp} \citep{Yee2017}, which characterises spectra of stars by comparing them with a library of high resolution spectra obtained with Keck/HIRES. 
\textsc{specmatch-emp} first shifts the input spectrum to the same wavelength scale as the templates, and then compares it to each star in the library to find the best matching  individual spectra. Linear combinations of the best matches are then used to create the best match to the input spectrum. Various stellar parameters for the star are then computed based on a weighted average of library parameters from the reference spectra. These properties are shown in Table \ref{tab:stellar_params}. 
The parameters calculated by \textsc{specmatch-emp} show that the star is a late K-dwarf with a mass of 0.73 \Msun.  Rather than adopting these as the final parameters for the primary star, we use them as priors for the spectral energy distribution (SED) fitting procedure detailed in \ref{ariadne}.  

It is interesting to note, that from this analysis we constrain the metallicity to a value of +0.1~dex, implying that \Nstar\ is a metal-rich star.  Even though in \citet{jenkins15} they show that the median brown dwarf host star metallicity is sub-solar, the distribution spans a wide range of values with a flat functional form, unlike the case of giant planets. Here. \Nstar\ adds another example to the metal rich population of brown dwarf host stars,and this time with a measured radius.

Additionally, we used this stacked spectrum to determine the projected stellar rotation velocity (v sin(i)). To do this, we fit synthesised spectra to the spectrum of \Nstar\ using iSpec \citep{Blanco2014}. We fit only for v sin(i), fixing the other values to those obtained from \textsc{specmatch-emp}. From this we obtain a value for vsin(i) of \vsini\, \kms\,.

\subsection{SED Fitting}\label{ariadne}
To obtain precise parameters for the host star, we performed a fit to the Spectral Energy Distribution. This was done using \textsc{ariadne} \citep{Vines2020}, which we describe in brief here. \textsc{ariadne} is a publicly available Python tool which fits catalogue photometry of stars from various sources (e.g., Gaia, TESS) to various atmospheric model grids. The specific grids used by \textsc{ariadne} are \texttt{Phoenix v2} \citep{Husser2013}, \texttt{BT-Settl}, \texttt{BT-Cond}, \texttt{BT-NextGen} \citep{Allard2012, Hauschildt99}, as well as the grids of \cite{Castelli2004}, and \cite{Kurucz1993}.

We create model SEDs by interpolating these grids in \teff\--\logg\--\feh\, space, with distance, radius and extinction in the $V$ band used as model parameters. An excess noise term was applied for each set of parameters to account for an underestimation of the uncertainties. These were normally distributed around zero with a variance of five times the size of the reported uncertainty. Priors for \teff\,, \logg\, and \feh\, and radius were applied based on the results from \textsc{specmatch-emp} (see Section \ref{specmatch}). Extinction was limited to the maximum line of sight value taken from the SFD galactic dust map (\citealt{Schlegel1998, Schlafly2011}). 

\begin{figure}
   \centering
	\includegraphics[width=\columnwidth]{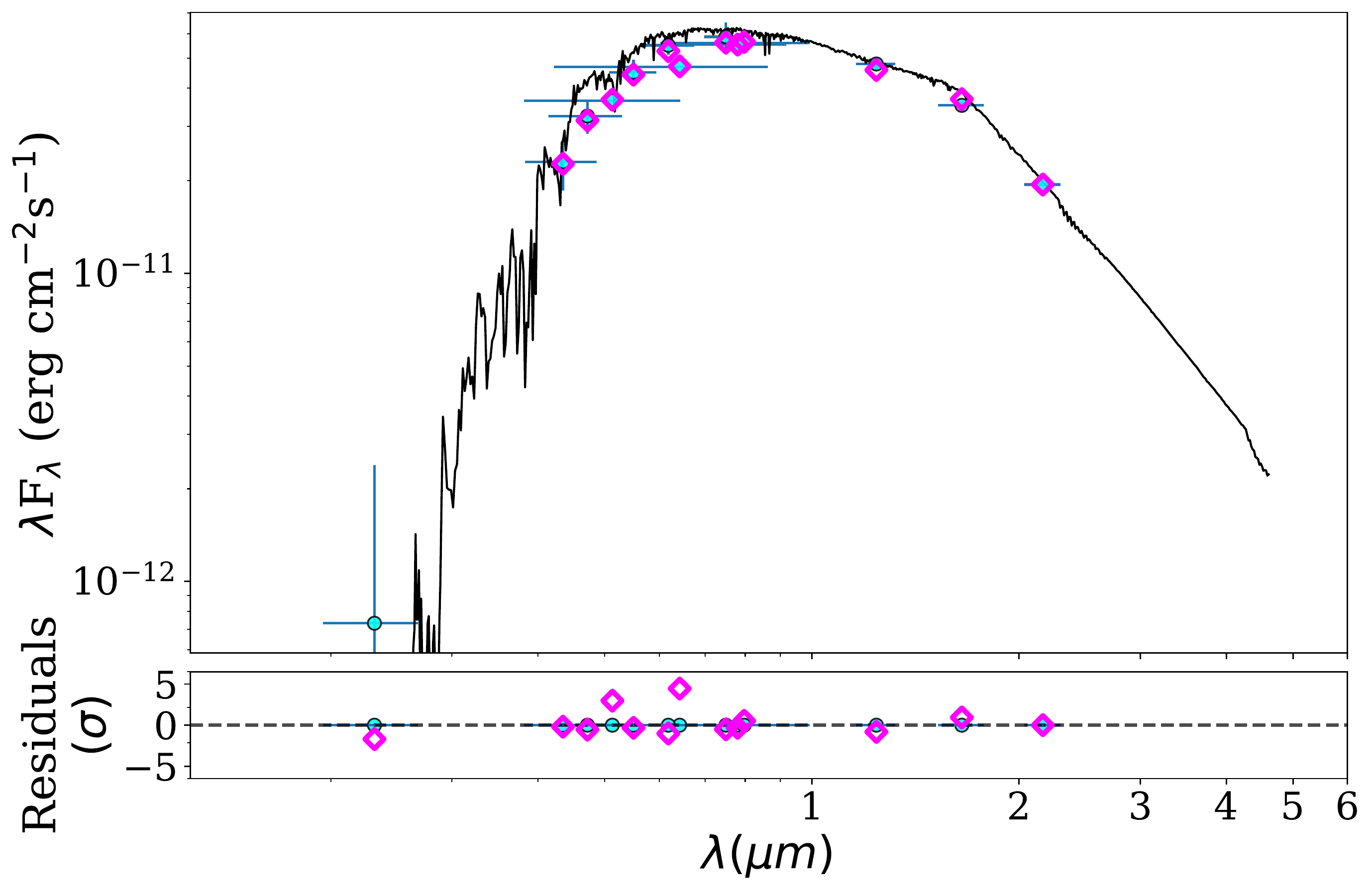}
   \caption{Upper Panel -- Spectral energy distribution of \Nstar\,. Blue points show the catalogue photometry for the system from a variety of sources, and the magenta diamonds are the synthetic photometry fit. The black line shows the best fitting model. Lower Panel -- Residuals to the SED fit, normalised to the photometry errors.}
   \label{fig:SEDFIT}
\end{figure}

The parameters from the SED were estimated using nested sampling performed using the python package \textsc{dynesty}. This was also used to calculate the Bayesian evidence for each of the individual models. For the fitted parameters, a weighted average is then computed using the relative probabilities of each of the fitted models. This Bayesian model averaging results in a remarkable degree of precision in these parameters when compared with using any one individual model SED fit. Finally, a mass estimate is calculate using a MIST isochrone \citep{Choi2016}. The results of this fitting are given in Table~\ref{tab:stellar_params}, and the fit to the SED is shown in Figure \ref{fig:SEDFIT}. Based on our SED fit and the classification established by \citet{Pecaut2013}, this makes the primary star a K3.5V dwarf. We adopt these parameters for the host star in our subsequent global fitting for this system 

\subsection{Global Modelling}\label{modelling}

\begin{figure}
   \centering
  \includegraphics[width=\columnwidth]{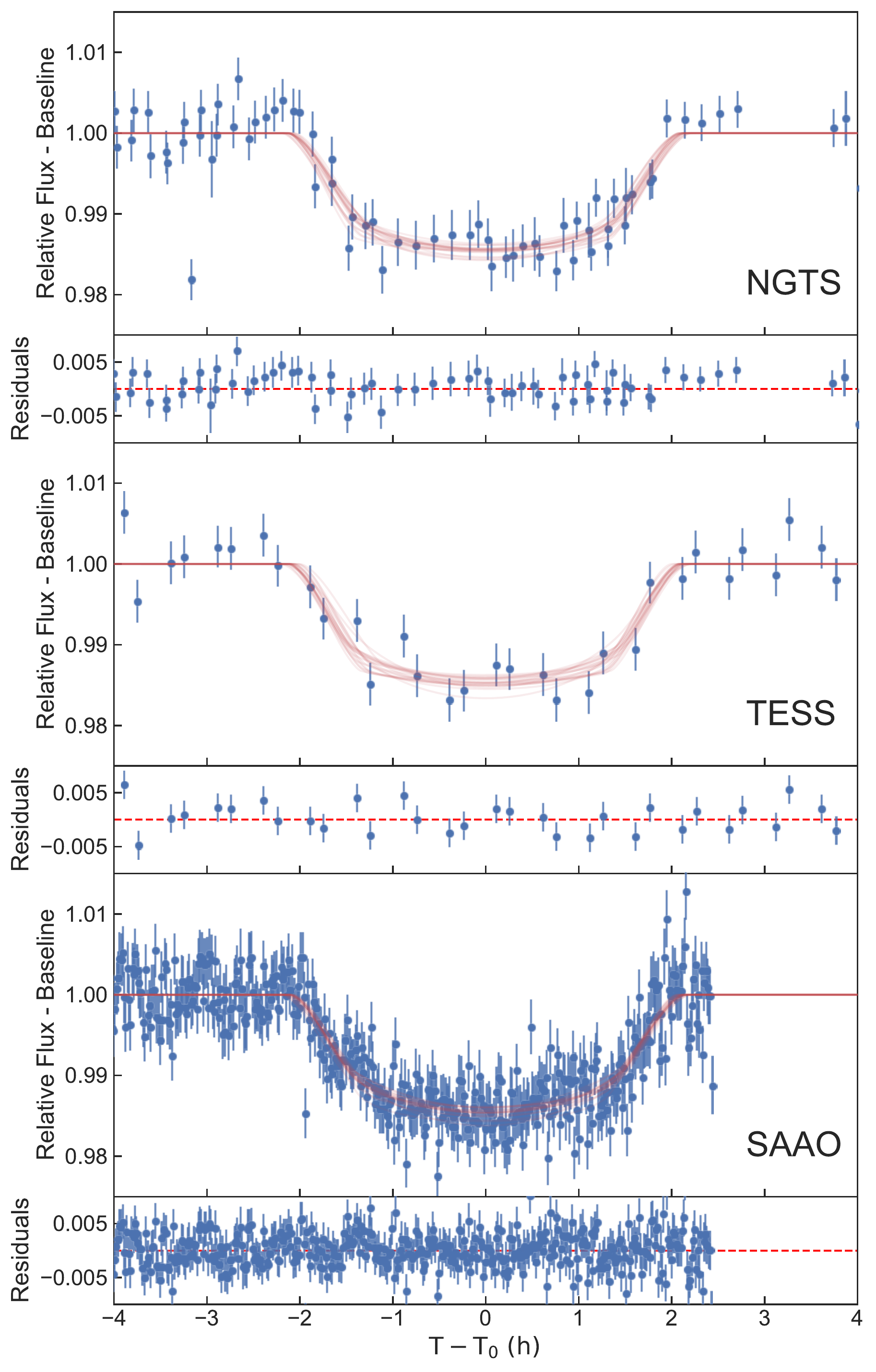}
   \caption{Data and model fits to the photometry from NGTS, TESS and SAAO-1m obtained using \textsc{allesfitter}. The NGTS data is binned to 10 minutes, whereas the TESS and SAAO data are unbinned. The red line shows the model fit derived by \textsc{allesfitter}, and the shaded region is the uncertainty in the model. }
   \label{fig:photfit}
\end{figure}

\begin{figure*}
   \centering
  \includegraphics[width=\textwidth,scale=0.1]{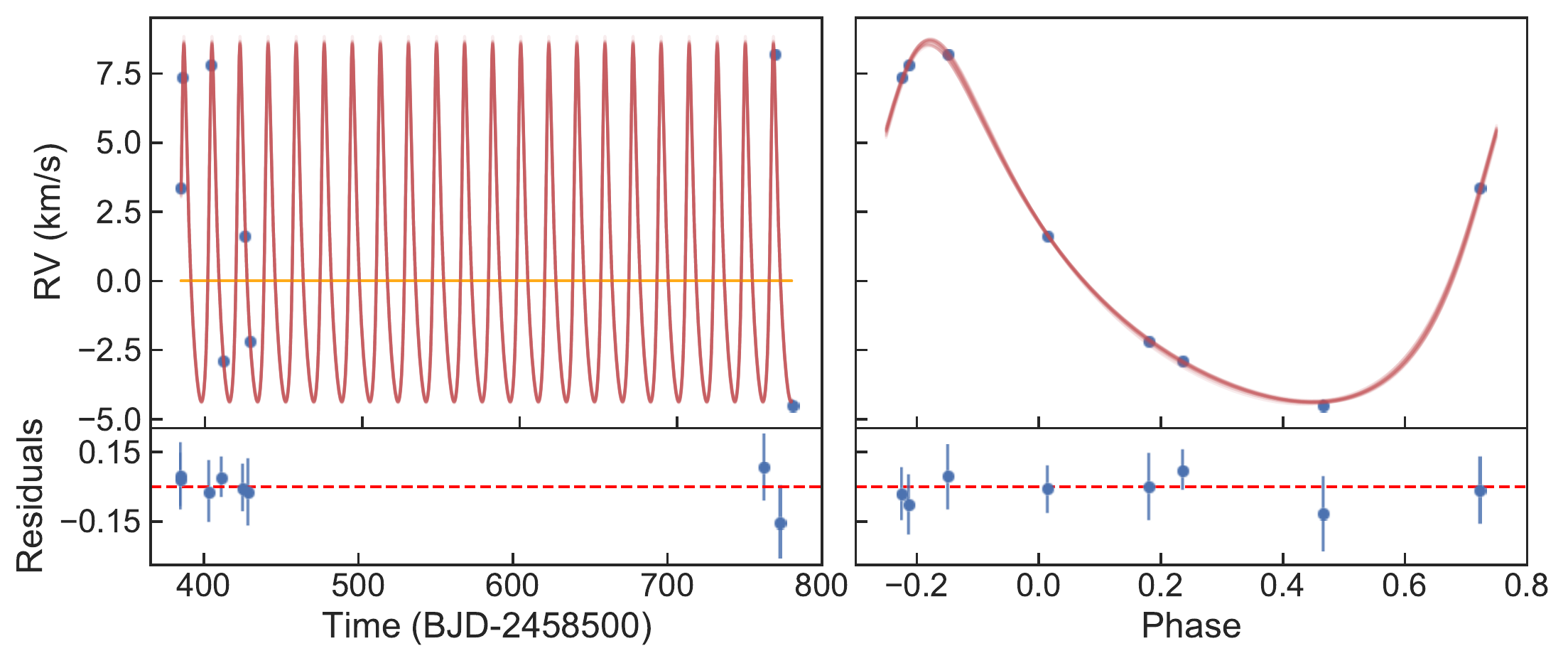}
   \caption{Data and model fits to the radial velocity measurements from CORALIE, obtained using \textsc{allesfitter}. The CORALIE radial velocities have had the systemic velocity of the \Nstar\ system subtracted, and a jitter term has been added in quadrature to the error bars. The red line shows the model fit derived by \textsc{allesfitter}, and the shaded region is the 1-sigma uncertainty in the model. The left panel shows the measurements in time series, and the right hand panel shows the radial velocities folded on the period determined by global modelling. }
   \label{fig:rvfit}
\end{figure*}

\begin{table}
 \caption[Model parameters]{Best fitting and derived parameters from the global modelling of \Nstar\ using \textsc{allesfitter}. The values are derived from the modal values of the posterior distribution, with 1-$\sigma$ uncertainties stated as errors. Note that the limb darkening coefficients are parameterised as in \citet{Kipping2013}} 
 \label{tab:globalfit}
 \resizebox{\columnwidth}{!}{
 \begin{tabular}{l c c c}
 \renewcommand{\arraystretch}{2.0}
 Parameter&   Symbol  &  Unit  & Value \\  
\hline 
 \multicolumn{4}{c}{\emph{Fitted Transit parameters}} \\ [1.0ex]
 Scaled BD and Star Radii  & $(\rm{R}_{\rm{*}}+\rm{R}_{\rm{BD}})/a$          & ---  & \Rsuma      \\ [1.0ex]
 Radius radio            & $\rm{R}_{\rm{BD}}/\rm{R}_{\rm{*}}$          & ---  & \RRatio      \\ [1.0ex]
 Cosine inclination      & $\cos i$                                   & ---  & \cosinc    \\ [1.0ex]
 Impact parameter        & b                                          & ---  & \impact       \\ [1.0ex]
 Epoch                   & T$_{0}$                                    & HJD  & \Epoch   \\ [1.0ex]
 Period                  & P                                          & days & \Period  \\ [1.0ex]
 $\sqrt{e}cos(\omega)$                      & $f_c$                                          & ---  & \eccone  \\ [1.0ex] 
 $\sqrt{e}sin(\omega)$                       & $f_s$                                          & ---  & \ecctwo  \\ [1.0ex] 
\multicolumn{4}{c}{\emph{Limb Darkening Parameters}} \\  [1.0ex]
NGTS LDC 1 &  $q_{1,NGTS}$   &  & \ldcngtsone       \\ [1.0ex]
NGTS LDC 2   &  $q_{2,NGTS}$   &  & \ldcngtstwo        \\ [1.0ex]
TESS LDC 1 &   $q_{1,TESS}$  &  & \ldctessone       \\ [1.0ex]
TESS LDC 2  &   $q_{2,TESS}$  &  & \ldctesstwo        \\ [1.0ex]
SAAO LDC 1 &    $q_{1,SAAO}$ &  & \ldcsaaoone       \\ [1.0ex]
SAAO LDC 2 &   $q_{2,SAAO}$  &  & \ldcsaaotwo        \\ [1.0ex]
\multicolumn{4}{c}{\emph{Radial Velocity Parameters}} \\  [1.0ex]
Systemic velocity  & V$_{\rm{sys}}$  & \kms & \sysvel       \\ [1.0ex]
RV semi-amplitude  & K               &\kms & \rvamp        \\ [1.0ex] 
\multicolumn{4}{c}{\emph{Derived brown dwarf parameters:}} \\  [1.0ex]
Brown dwarf mass              &  $M_{BD}$    &\Mjup    & \BDMass   \\ [1.0ex]
Brown dwarf radius            &  $R_{BD}$      &\Rjup    & \BDRad  \\ [1.0ex]
Brown dwarf density           &  $\rho_{BD}$    &   g cm\textsuperscript{-3}     & \BDDens \\ [1.0ex]
Semi-major axis          &      a     & AU      & \SMAAU     \\ [1.0ex]
Scaled Semi-major axis          &  $a/R_{*}$         & ---     & \SMAscaled    \\ [1.0ex]
Inclination        &     i      & Degrees     & \incl    \\ [1.0ex]
Orbital Eccentricity         &     e      & ---     & \ecc    \\ [1.0ex]
Argument of Periastron         &     $\omega$      & Degrees     & \periastron    \\ [1.0ex]
Equilibrium Temp.        &     $T_{eq,BD}$        & K       & \eqtemp \\[1.0ex]
Transit Duration       &    $t_{trans}$       & hours       & \transitwidth \\[1.0ex]
\\ [1.0ex]
 \noalign{\smallskip} \noalign{\smallskip}  
 \hline  
 \end{tabular}  
 }
 \end{table} 
 
To determine orbital parameters for the system we jointly modelled the photometry and radial velocities using \textsc{allesfitter} (\citealt{ allesfitter-code,allesfitter-paper}). \textsc{allesfitter} is a publicly available Python code for performing global modelling of photometric and radial velocity data. The software acts as a wrapper for a number of well-used packages, in particular, we utilise the eclipsing binary light curve modeller \textsc{ellc} \citep{Maxted2016} and the Markov Chain Monte Carlo (MCMC) sampler \textsc{emcee} \citep{Foreman-Mackey2013} to simultaneously model the NGTS, TESS, SAAO and CORALIE data. Prior to running the MCMC we perform some simple preparation of the data. The raw lightcurves were normalised to a baseline of 1 by using the median out of transit flux. We then binned the NGTS data to 10 minutes, to reduce computational time. The TESS and SAAO data remained unbinned. 

We chose to start the walkers in a region of parameter space that gave a reasonable initial fit to the data. We note that whilst starting the walkers in a random position does not preclude the ability to obtain a good fit to the data, it does increase the burn in time required to do so. Each walker was given a starting position normally distributed around the values we found to give an initial fit. We used values from \textsc{orion} for the ephemeris of the system, and obtained values for the stellar radius ratio, scaled primary star radius and inclination from an initial fit the the NGTS data alone performed by \textsc{orion}. Equally, starting positions for the radial velocity parameters were determined by fitting these data alone. We also incorporated a radial velocity jitter term added in quadrature to account for any effects of stellar variability in the RVs, as well as normalisation offsets and systematic errors for the lightcurves. We knew from the radial velocity measurements that there was a significant level of eccentricity in the system, but for the fitting we started the walkers at zero eccentricity and allowed the sampler to determine it. We note that for fitting purposes, \textsc{allesfitter} parameterises eccentricity $e$, relative to the argument of perisastron, $\omega$ , using terms $\sqrt{e}cos(\omega)$ and $\sqrt{e}sin(\omega)$. Finally, we also determined limb darkening parameters using the \textsc{ldtk} package \citep{Parvianen2015} for each photometric filter used, adopting limb darkening laws from \citet{Kipping2013} in the fit. 

We ran \textsc{allesfitter} with 100 walkers going for 80000 steps. We found this to be more than sufficient for the MCMC to converge to a solution, ensuring that the chains were at least 30 times the autocorrelation length of each parameter. 10000 of these steps were discarded as burn in and not used when analysing the results. The modal values of the posterior distributions for each parameter were adopted as the most probable values, with the 1-$\sigma$ (68.3 percent) confidence intervals taken as an estimate of uncertainty. The fitted parameters were then used to derive additional parameters for the system.

This global modelling reveals the companion in this system is a high mass brown dwarf, with a mass of \BDMass\,\Mjup\, and radius of \BDRad\,\Rjup\. The system is also notable for having a reasonably long period of \Period\, days, and a high level of eccentricity \ecc. We note that this period makes \Nstar\ the longest period transiting brown dwarf to be initially discovered using purely ground-based photometry, highlighting the value of the long observing baselines used by NGTS.
The full list of fitted and derived parameters for the system is given in Table \ref{tab:globalfit}. The model fits to each set of data are shown in Figures \ref{fig:photfit} and \ref{fig:rvfit}.

\section{Discussion}

\subsection{Mass-Radius Relation}\label{massradius}
In Section \ref{modelling} we determined that the companion to \Nstar\, is a high mass brown dwarf. This adds to the small, but growing number of objects being discovered in the so-called "brown dwarf desert". With direct measurements of the brown dwarf's mass and radius  
we can make a comparison with both the known population of transiting brown dwarfs, as well as with predictions from evolutionary models. In Figure \ref{fig:mass-radius}, we compare the mass and radius of \Nstarb\ to the known brown dwarf companions around main sequence stars from \citet{Carmichael2020}. We see that with a mass of \BDMass \Mjup, \Nstarb\ lies at the upper end of the brown dwarf mass distribution, close to the substellar boundary. 

When we compare the system to isochrones from  \citet{Marley2018}, we see that the system agrees well with the 0.4 Gyr model.  This is in contrast to the age determination from the SED fit performed by \textsc{ariadne}, which predicted a much older system age of $8.5 ^{+3.2}_{-6.0}$ Gyr. This leaves us with two potential scenarios. It is possible that the system is indeed young, despite the age estimates from the SED. Alternatively, we may assume that the age from the SED fit is correct and the brown dwarf's radius has been inflated by some mechanism, making it appear younger in mass-radius space.

\subsubsection{Possible youth of the system}\label{youth}

An important parameter to know for any discovered transiting brown dwarf is the system age. To gain some estimate of this, we compare the mass and radius of \Nstarb\ to evolutionary models from, for a variety of ages. We follow the method of \citet{Gillen2020}, first interpolating between the models in order to compute a finer grid of model predictions and then comparing our global modelling results for the mass and radius of the system to this grid to compute an age estimate. For this, we use the Sonora brown dwarf models of \citet{Marley2018}. Doing so yields an estimate for the age of the system of $0.46 ^{+0.26}_{-0.15}$ Gyr, this is not consistent with the determined age from \textsc{ariadne}, which predicts an older system.

Despite the small number of transiting brown dwarfs yet discovered, there are a reasonable number of young systems. These are often found in clusters, from which the age can be accurately determined (\citealt{Gillen2017, Beatty2018, David2019}). However there are known brown dwarfs orbiting young stars not associated with clusters, for example NGTS-7Ab \citep{Jackman2019}, with an age of just 55 Myr. Discoveries of brown dwarfs at these ages is important, as they will be undergoing gravitational contraction at a faster rate than at later ages (as seen by the large gaps between isochrones in Figure \ref{fig:mass-radius}). This allows for some of the most powerful tests of brown dwarf evolutionary models. 

This however does not explain the discrepancy seen between the age determined from the brown dwarf mass \& radius, and that found in the SED fit. We note that the age determined from the SED by \textsc{ariadne} is poorly constrained, and is consistent with our inferred brown dwarf age at the 2$\sigma$ level. To try and confirm or refute this age measurement, we checked for additional youth indicators. We examined the CORALIE spectrum for signs of Lithium absorption, which is typically an indicator of youth, however we find no such absorption in the spectrum. The star also has a relatively low vsin(i), which would also not suggest youth, and is not part of a known moving group. Thus the only evidence that the brown dwarf is young is its unusually large radius. This means that we cannot conclusively say that this is indeed a young system, as the radius of the brown dwarf could have been increased by some other means.

\subsubsection{Inflated Brown Dwarf Radii}
\begin{figure*}
	\includegraphics[width=\textwidth]{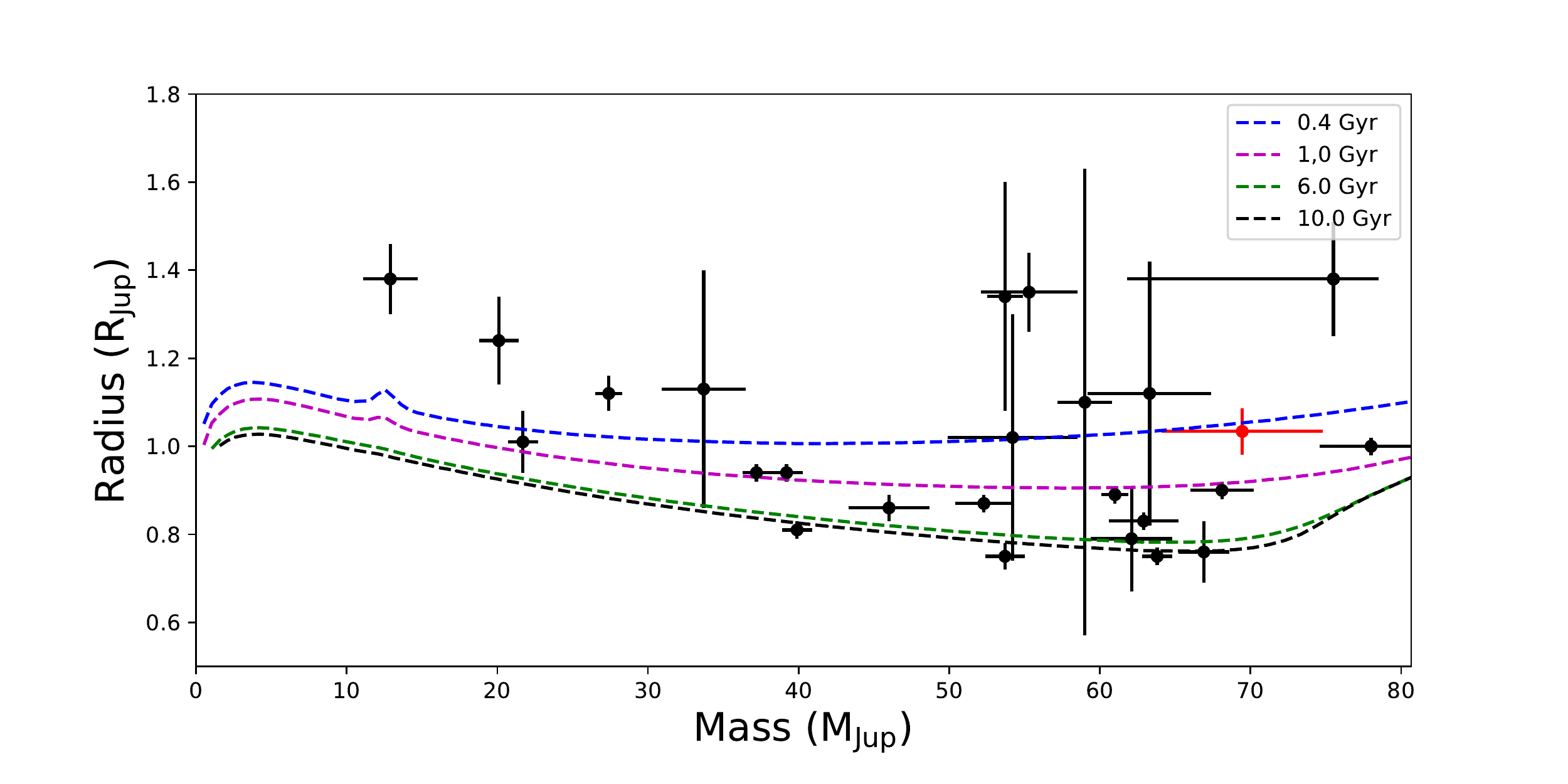}
    \caption{Mass-Radius relation for known brown dwarfs around Main Sequence stars from \citet{Carmichael2020}. \Nstarb\ is plotted in red. Sonora model isochrones from \citet{Marley2018} for ages of 0.4, 1, 6 and 10 Gyr are plotted for comparison. Note that RIK 72b \citep{David2019} is not shown due to its inflated radius of 3.1\Rjup, nor is the brown dwarf binary system 2M0535-05a \citep{Stassunn2006}}
   \label{fig:mass-radius}
\end{figure*}

 It should be noted however, that the models used in Figure \ref{fig:mass-radius} are for isolated brown dwarfs, whereas here we have a brown dwarf with a close stellar companion. Therefore, it is possible that the system is indeed the age determined from the SED and  \Nstarb\ is inflated relative to model predictions. There are brown dwarfs with unusually large radii, however these are usually associated with being strongly irradiated (e.g \citealt{Zhou2019, Siverd2012}), although this is not always the case (e.g \citealt{Csizmadia2015}). 

\begin{figure*}
	\includegraphics[width=\textwidth]{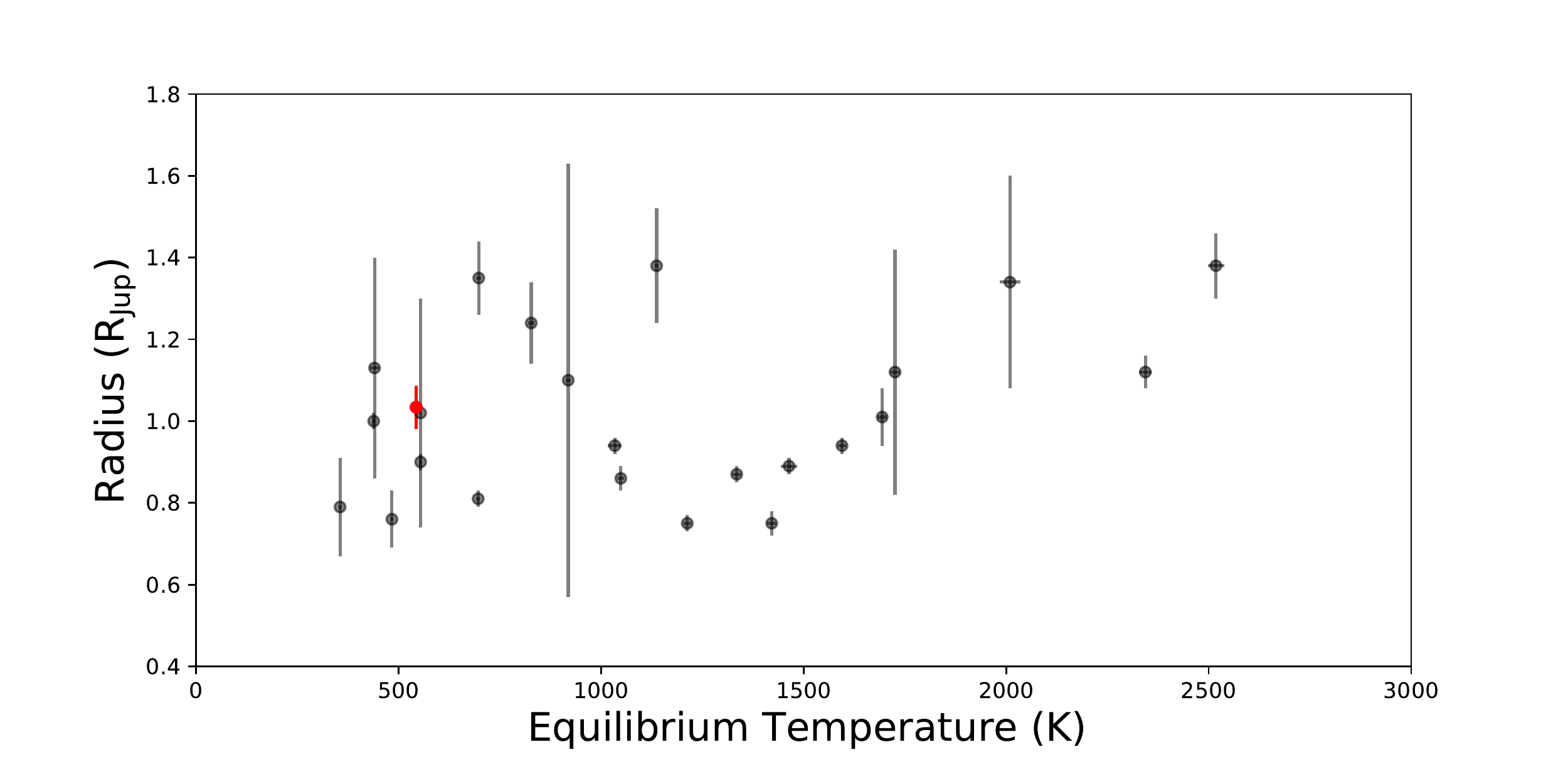}
    \caption{Relationship between equilibrium temperature and radius for known brown dwarfs around main sequence stars from \citet{Carmichael2020}. \Nstarb\ is plotted in red.  We use the effective temperature of the host star (if reported) and the relation in \citet{Mendez2017} to calculate the time-averaged equilibrium temperature of the brown dwarf. For all systems we assume a Bond albedo of zero. Note that RIK 72b \citep{David2019} is not shown due to its inflated radius of 3.1\Rjup, nor is the brown dwarf binary system 2M0535-05a \citep{Stassunn2006}}
   \label{fig:eq-temp}
\end{figure*}

\cite{Bouchy2011} showed that unlike in the case of hot Jupiters, where stellar irradiation can play a large effect in inflating the radii of the companion \citep{Sestovic2018}, the effect is much smaller for brown dwarfs, particularly at the higher mass end. 
It is also worth noting that the brown dwarf they were analysing (CoRoT-15b) has an equilibrium temperature almost twice that of \Nstarb\,. Hence it is even less likely to have have an effect on the radius of this system. In Figure \ref{fig:eq-temp} we plot the relationship between the radius and equilibrium temperature for known brown dwarfs. We see that although there is perhaps a slight upwards trend in radius with increasing temperature it is not noticeably significant. When compared to brown dwarfs with similar equilibrium temperatures, we see that the radius of \Nstarb\ is not an outlier, and is well within the observed scatter. If \Nstarb\ is truly an inflated brown dwarf then there must be some other means to have increased its radius other than purely irradiation from its host star. 

It is important to note however that the orbit of \Nstarb\ is highly eccentric, and thus the distance between the brown dwarf and its host star varies significantly. Based on the measured eccentricity and semi-major axis, we determine a periastron distance of $0.0817$ AU, which is not too dissimilar to that of a hot Jupiter. Therefore it is possible that the increased irradiation experienced by the brown dwarf at periastron could act to inflate its radius. This eccentric orbit could also act to inflate the brown dwarf due to tidal effects exerted on it by the host star, as has been demonstrated for gas giant exoplanets \citep{Millholland2020}. However given the long period of the system, and the high mass of the brown dwarf, we determine it to be unlikely that these effects would transfer enough heat into the brown dwarf to inflate it.


\cite{Casewell2020} consider the white-dwarf brown-dwarf binary NLTT5036, which also shows signs of inflation. They show that regardless of the level or irradiation experienced, low mass brown-dwarfs (M$<$35\Mjup) are able to be inflated when heated by their host star. However, higher mass brown-dwarfs are much harder to inflate, with the majority of them showing no inflation. The temperature of the host star also plays an important role here; hotter stars will emit more ultraviolet radiation whereas cooler stars will emit more in the infrared. \Nstar\,, a mid K-dwarf, has an effective temperature of 4890 K which would leave it at the cooler end of the distribution of transiting brown dwarf host stars \citep{Carmichael2020}, with no significant amount of either UV or IR radiation. 

Stellar activity may also affect the ability of a brown-dwarf to be inflated by its host star. CoRoT-15b \citep{Bouchy2011} and CoRoT-33b \citep{Csizmadia2015} both show signs of being inflated, and are noted to be orbiting active host stars. This may suggest that brown-dwarf inflation is caused by a similar mechanism to M-dwarfs \citep{Stelzer2013}. However these two brown dwarfs both have large radii uncertainties (27\% and 48\% for CoRoT-15b and 33b respectively) so it is difficult to categorically say whether or not these systems are actually inflated. We note that \Nstar\ does not show any signs of significant stellar activity. In almost 150 nights of NGTS data we do not detect any obvious flares, and the CORALIE spectra do not show any signs of magnetic activity, such as H$\alpha$ emission.  However to accurately characterise the stars magnetic field would require further investigation.

Another important factor to consider is metallicity. For a fixed mass and age, the radii of brown dwarfs increase with increasing metallicity. \citet{Burrows2011} show that a change from $+$0.0~dex to $+$0.5~dex in the metallicity of the brown dwarf can result in an increase in radius of as much as 0.1 \Rjup. Our analysis in Sections \ref{specmatch} and \ref{ariadne} suggest a small, but significant host star metallicity of around 0.1~dex. This may explain some of the discrepancy between model and measurement we see for \Nstarb\,, but is unlikely to increase the radius by enough given the disagreement between our measured and predicted ages. They also show that differences between clear and cloudy brown dwarf models can cause deviations in radius by about 0.05 \Rjup, but this too is likely not substantial enough to account for the discrepancy seen in our measurements.

\subsection{Eccentricity and Tidal Circularisation}
\begin{figure}
	\includegraphics[width=\columnwidth]{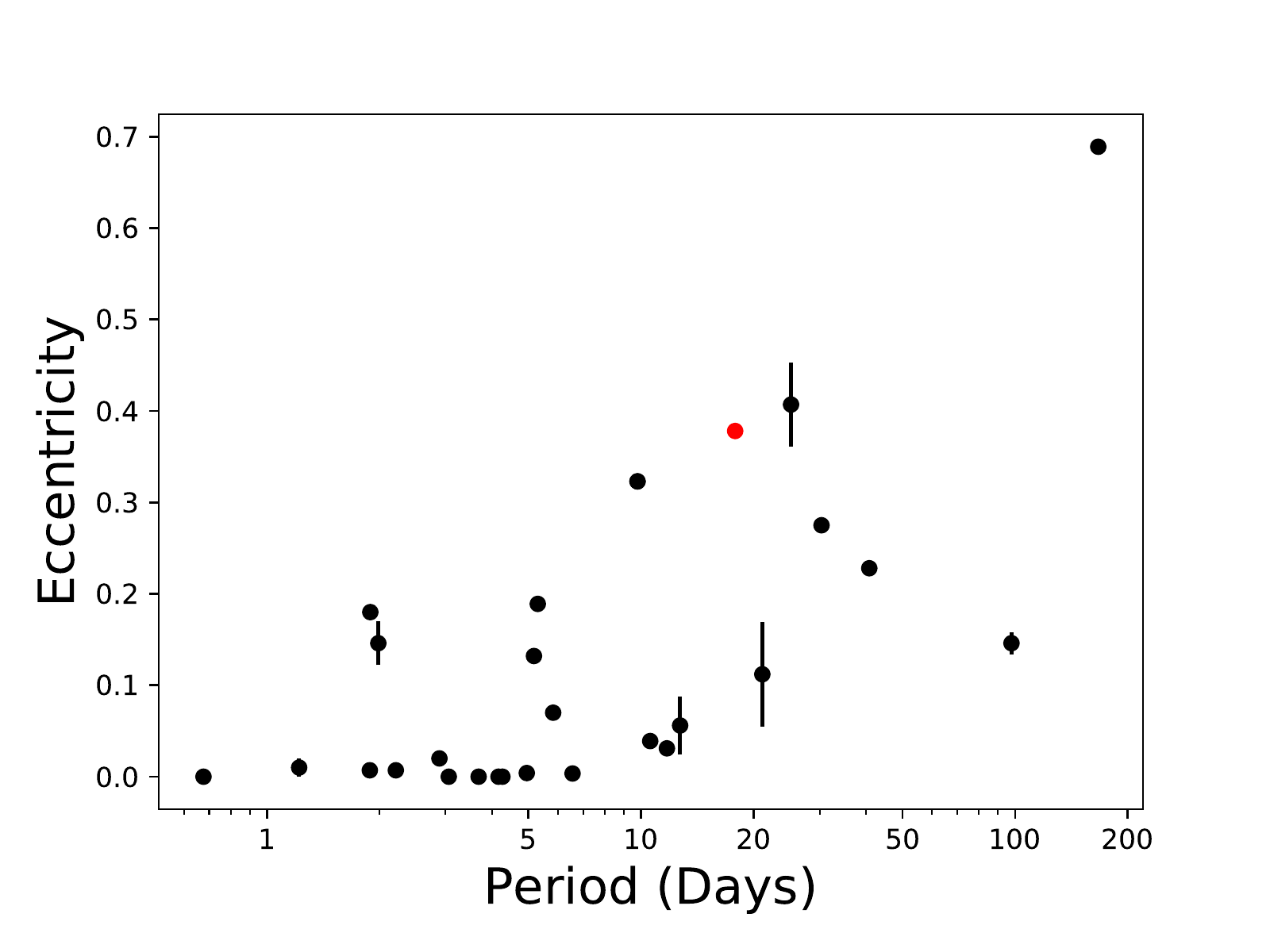}
    \caption{Eccentricity against period for known transiting brown dwarfs orbiting main sequence stars from \citet{Carmichael2020}. \Nstarb\ is indicated in red.}
   \label{fig:ecc-period}
\end{figure}

\Nstarb\ has a relatively large orbital eccentricity of \ecc\, (Section \ref{modelling}). This is not necessarily unusual for transiting brown dwarfs. Of the 27 transiting brown dwarfs listed by \citet{Carmichael2020}, 10 have eccentricity greater than 0.1. In Figure \ref{fig:ecc-period}, we show the distribution of eccentricity as a function of orbital period for the population of transiting brown dwarfs. Only two known systems have a higher eccentricity than \Nstarb\,, these are TOI-811b and KOI-415b (\citealt{Carmichael2020,Moutou2013}) with orbital eccentricities of $0.4072\pm 0.046$ and $0.689\pm 0.001$ respectively. However with orbital periods of 25.2 and 166.8 days, these systems are both longer period than \Nstarb\, and there is no known transiting brown dwarf with a shorter orbital period that has a higher degree of eccentricity. 

For a system with a semi-major axis of just \SMAAU\, AU, this is a reasonably high level of eccentricity. It is known that over the course of evolution of such small separation systems, the tidal effect of the host star acts to circularise the orbit. This was a process first applied to binary stars \citep{Zahn1989}, but is equally applicable to brown dwarf and hot Jupiter systems \citep{Rasio1996}. We can apply this theory to \Nstarb\,, to evaluate whether we would expect the orbit to have circularised, given the period and masses of the system. The majority of this orbital circularisation is expected to occur early in the lifetime of the system, therefore if we determine a short orbital circularisation timescale for \Nstarb\ it could be an indicator that the system is indeed young as suggested in section \ref{youth}.

We follow the method of \citet{Carmichael2020b}, based on the theoretical framework of \citet{Jackson2008}. Here the orbital circularisation timescales for the host star and brown dwarf are defined as

\begin{equation}
\centering
    \frac{1}{\tau_{circ,*}}=\frac{171}{16}\sqrt{\frac{G}{M_*}}\frac{R^5_* M_{BD}}{Q_*}a^{\frac{-13}{2}} \; , \hspace{2mm}{\rm and}
\end{equation}{}

\begin{equation}
\centering
    \frac{1}{\tau_{circ,BD}}=\frac{63}{4}\frac{\sqrt{G M^3_*} R^5_{BD}}{Q_{BD}M_{BD}}    a^{\frac{-13}{2}} \; ,
\end{equation}{}

\noindent where $a$ is the semi-major axis of the system, and $Q_*$ and $Q_{BD}$ are the tidal quality factors for the star and brown dwarf respectively. From this, the circularisation timescale of the system is defined as 

\begin{equation}
\centering
    \frac{1}{\tau_{e}}= \frac{1}{\tau_{circ,*}} +  \frac{1}{\tau_{circ,BD}} \; .
\end{equation}{}

\noindent From this we can determine the circularisation timescale for the system. We note that this is typically applied to systems with lower eccentricities and shorter orbital periods than \Nstarb\,. However, it is still a useful method for estimating whether or not we would expect a system of this type to have circularised in its lifetime. The tidal quality factors are rather poorly constrained in the literature, so as in \citet{Carmichael2020b} we adopt a lower bound on $Q_{BD}$ of $10^{4.5}$ based on the work of \citet{Beatty2018} on CWW 89Ab. We place a lower bound of $10^5$ on $Q_*$ based on previous studies of binary stars (\citealt{Meibom2005, Milliman2014}).

In Table~\ref{tab:tides} we show the circularisation timescale for various combinations of both stellar and brown dwarf tidal quality factors. We find that there is no combination of tidal quality factors that provide a circularisation timescale of less than 18 Gyr, given the lower bounds we have placed. In fact, almost all combinations give a large circularisation timescale. Unless the tidal quality factors are substantially smaller than our lowest estimates (by at least an order of magnitude), we conclude that we would not expect the system to have been tidally circularised, and thus finding a brown dwarf of this orbital period and eccentricity is not unusual.

\addtolength{\tabcolsep}{22pt} 
\begin{table}
	\centering
	\caption{Tidal circularisation timescale for \Nstarb\ based on the tidal circularisation model of \citet{Jackson2008}, for a combination of reasonable stellar and brown dwarf tidal quality factors. For no realistic combination of tidal quality factors do we find a circularisation timescale that is reasonable.}
	\label{tab:tides}
\begin{tabular*}{\columnwidth}{ccc}
\hline

     $Q_*$        & $Q_{BD}$        &   $\tau_e (Gyr)$     \\ \hline
     \hline
    $10^{5}$           &      $10^{4.5}$       &     $18.39 $         \\
    $10^{6}$            &     $10^{4.5}$        &     $169.10$          \\
    $10^{7}$            &     $10^{4.5}$        &     $934.54$          \\
    $10^{5}$            &     $10^{5}$          &     $18.52  $        \\
    $10^{6}$            &     $10^{5}$          &     $180.18  $        \\
    $10^{5}$            &     $10^{6}$          &      $18.58   $      \\
    $10^{6}$            &     $10^{6}$          &       $185.23  $      \\
    $10^{7}$            &     $10^{6}$          &       $1801.78  $      \\

     \hline
\end{tabular*}
\end{table}
\addtolength{\tabcolsep}{-22pt}

\subsection{Future observing prospects}

\subsubsection{Secondary Eclipse of \Nstarb\,}
It would be particularly valuable to our understanding on \Nstarb\ if we were able to detect a secondary eclipse of the brown dwarf. Doing so would allow for direct determination of the brown dwarf temperature (as shown in \citet{Jackman2019}). However, due to the significant difference in luminosity between the host star and the fainter brown dwarf, making such a detection is difficult. 

\begin{figure}
	\includegraphics[width=\columnwidth]{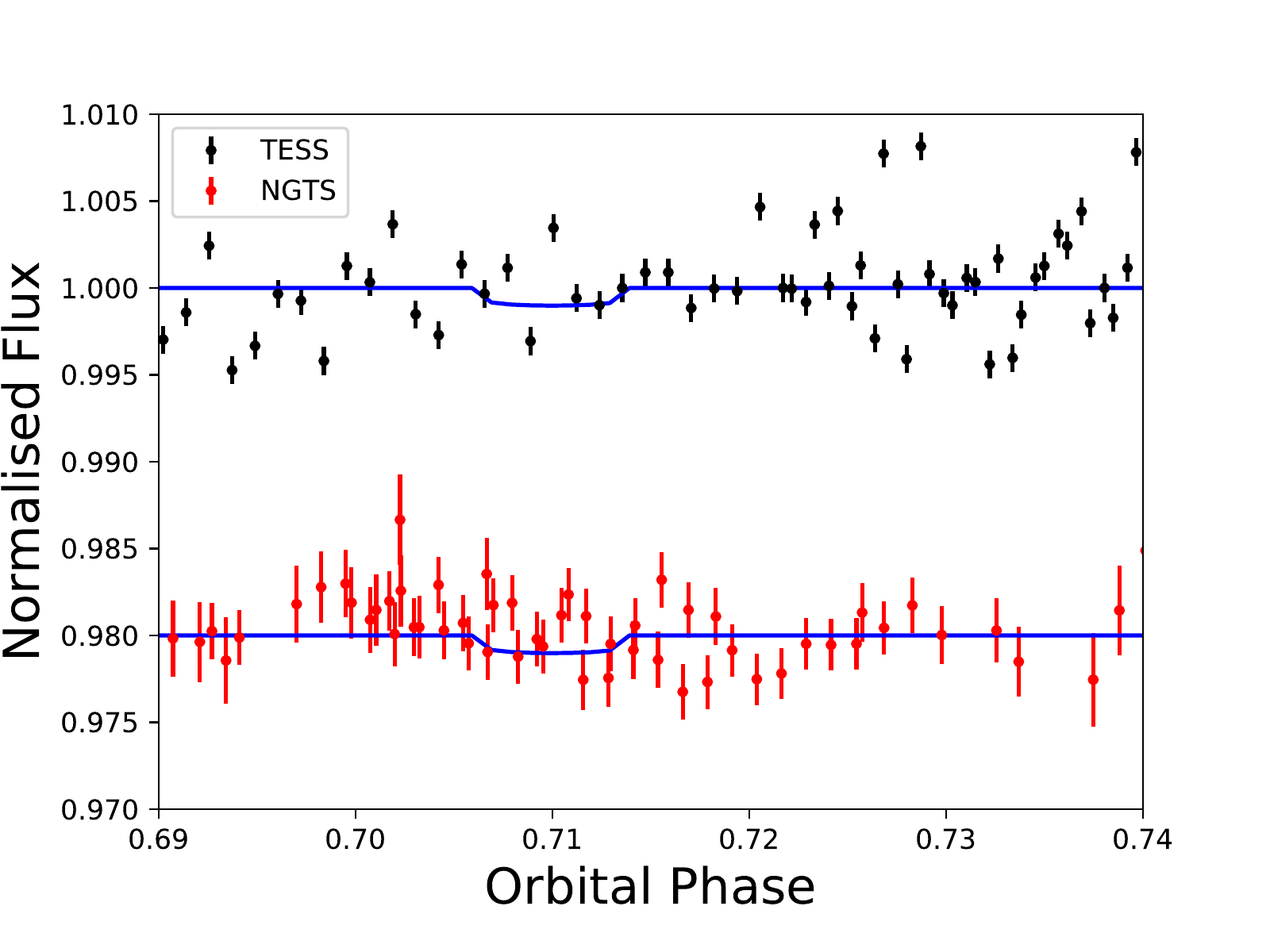}
    \caption{NGTS and TESS photometry for \Nstar\ around the expected phase of the secondary eclipse of the system. The NGTS data is binned to 30 minutes for comparison with the TESS full frame image data. Over plotted in blue is a model of the maximum depth secondary eclipse for comparison. No obvious secondary eclipse is seen in either of the lightcurves.}
   \label{fig:secondary-eclipse}
\end{figure}

For an eccentricity of \ecc\  (assuming the argument of periastron in table \ref{tab:globalfit}), we would expect to see a secondary eclipse at around phase \secphase\,. Both NGTS and TESS observations have coverage at this phase, so we examined their lightcurves for evidence of this secondary eclipse (Figure \ref{fig:secondary-eclipse}). Despite the level of scatter in both lightcurves being very small ($<$0.1\%) for an object this faint, there is no obvious sign of a secondary eclipse seen in either of the lightcurves. Indeed we searched a large range of orbital phase to account for any error in our calculation of orbital eccentricity, and we see no evidence for a secondary eclipse at any point. We note that for some objects in eccentric orbits you do not see a secondary eclipse at all, due to the system configuration and inclination. However this is not the case for \Nstarb\, where we would expect to see a secondary if the object had a high enough surface brightness, based on our modelling.

Based on the atmospheric models of \citet{Baraffe2003}, an isolated 0.07 \Msun\ brown dwarf has an effective temperature of 1626~K at 10~Gyr. If the system is young as speculated in Section \ref{massradius}, then the temperature could instead be as high as 2335~K. This leads to predicted secondary eclipse depths of $\approx$0.008\% and $\approx$0.09\% respectively. This is within the scatter of both the NGTS and TESS lightcurves, and suggests that a non-detection is not unexpected. Due to the faint nature of this system, the detection of such a shallow secondary eclipse would be extremely challenging, even if the system is younger and more luminous than we expect. Based on this lack of secondary eclipse detection, we can place a tentative upper limit of $\sim$  2800 K on the brown dwarf temperature. For a temperature greater than this we would expect the secondary eclipse to be large enough that it would be visible in the lightcurves. This places a lower limit on the age from the \citet{Baraffe2003} models of 0.1 Gyr, so is not sufficient to constrain confirm or refute the suggested youth of the system, i.e even if the brown dwarf is as young as suggested by its radius, we still would not expect to detect a secondary eclipse.  

\subsubsection{Spin-orbit angle}
Transiting brown dwarfs present an interesting avenue to help understand the distribution seen in spin orbit angle seen for exoplanets and low mass stars. By measuring the Rossiter-McLaughlin effect, it has been shown that hot Jupiters can show a wide range of spin-orbit angles, with some even showing retrograde orbits \citep{Queloz2010}. However, similar studies of this effect for low mass stars has shown that these systems do tend to be aligned \citep{Triaud2013}. This is likely due to the fact that hot Jupiters arrive into their short period orbits via dynamical interactions that force them into highly eccentric orbits that then circularise - leading to misalignment. 

Brown dwarfs bridge the gap between planets and stars, so measuring the spin orbit angle of transiting brown dwarfs allows for insight into what may cause the misalignment seen in planetary systems. For this system in particular, detection of a misaligned orbit may be some indication that the large eccentricity may not be primordial, and has been caused interactions with a third body via Lidov-Kozai cycles (\citealt{Lidov1962,Kozai1962}). This would provide valuable insight into how a system such as this has formed and evolved.

The semi-amplitude of the Rossiiter-McLaughlin effect scales approximately with planet size and stellar rotational velocity, in the following relation \citep{Triaud2018}

\begin{equation}
    A_{RM} \approx \frac{2}{3}D vsin(i) \sqrt{1-b^2}
\end{equation}

Where $D$ is the transit depth $(\frac{R_{BD}}{R_*})^2$, $vsin(i)$ is the projected stellar rotation velocity, and $b$ is the impact parameter. From analysis of the CORALIE spectra, we calculate  $vsin(i)$ = \vsini\,\kms\, for \Nstar\,. This implies a Rossiter-McLaughlin amplitude of \RMAmp\,\ms\,. A signal of this magnitude could be detected easily with an instrument such as ESPRESSO \citep{Pepe2021}.

\section{Conclusions}
We report the discovery of a high mass brown dwarf companion on a P=\Period\,day eccentric orbit around a main sequence K type star. \Nstarb\ is a brown dwarf with a mass of \BDMass, placing it at the high end of the brown dwarf mass distribution. When compared to evolutionary models, the system is consistent with having an age of around 0.5 Gyr, although this is not consistent with age measurements from our spectroscopic or SED analysis, suggesting that the brown dwarf may be inflated due to interactions with its host star.

The system also has a highly eccentric orbit, with only 2 transiting brown dwarf systems being more eccentric. There are no shorter period, more eccentric transiting brown dwarfs known. When examined using established tidal circularisation theory, we find that the system has a long enough period that we would not expect it to have circularised in any reasonable time period.

\Nstarb\ is the 29\textsuperscript{th} transiting brown dwarf to be discovered, adding to the population of companions to main sequence stars known as the brown dwarf desert. With the continuing survey of the TESS mission it is quite possible that there will be many more additions to this once sparse region of parameter space in the years to come. 

\section*{Acknowledgements}
Based on data collected under the NGTS project at the ESO La Silla Paranal Observatory. The NGTS facility is operated by the consortium institutes with support from the UK Science and Technology Facilities Council (STFC) under projects ST/M001962/1 and ST/S002642/1. 

This paper includes data collected by the TESS mission. Funding for the TESS mission is provided by the NASA Explorer Program. 

This paper uses observations made at the South African Astronomical Observatory (SAAO). 

JA is supported by an STFC studentship. This work has been carried out within the framework of the National Centre of Competence in Research PlanetS supported by the Swiss National Science Foundation.  
JSJ acknowledges support by FONDECYT grant 1201371, and partial support from CONICYT project Basal AFB-170002.
MNG acknowledges support from MIT's Kavli Institute as a Juan Carlos Torres Fellow.
EG gratefully acknowledges support from the David and Claudia Harding Foundation in the form of a Winton Exoplanet Fellowship.

\section*{Data Availability}
The data underlying this article will be shared on reasonable request to the corresponding author.




\bibliographystyle{mnras}
\bibliography{ref} 




\appendix


\bsp	
\label{lastpage}
\end{document}